\documentstyle[12pt]{article}
\input{psfig}
\long\def\comment#1{}
\begin{document}
\title{A Formula for the Rotation Periods of the Planets \& Asteroids}

\author{Subhash Kak\\
Department of Electrical \& Computer Engineering\\
Louisiana State University\\
Baton Rouge, LA 70803-5901, USA\\
FAX: 225.388.5200; Email: {\tt kak@ee.lsu.edu}}

\maketitle

\begin{abstract}
The note presents a formula for the prediction of the
rotation periods of the planets and asteroids. This
formula, which is like the Titius-Bode law, gives
a good agreement with the rotation periods of most planets,
shows that Venus is retrograde, and that there must be
five objects between Mars and Jupiter. This formula may
be of some relevance in understanding the dynamics of the
early solar system.

\end{abstract}

It is generally
believed that the Titius-Bode relationship between the distance of the planets
from the sun may have some significance regarding the formation
of the solar system.
If there is
a similar simple pattern defining the
rotation periods of the planets then that may also provide
clues regarding the dynamics of the early solar system.
In this note I present
a simple relationship that is in good agreement with
the rotation period information of the superior planets,
and it indicates that Venus has retrograde rotation although
it does not give an accurate value of the rotation of this
planet or Mercury.

Instead of considering the rotation periods directly, I
consider $M = \frac{d}{p}$, where $d$ is the distance
from the sun in astronomical units (AU) and $p$ is the
rotation period in days. $M$ is a measure of the
relative speed of the sun across the horizon. The
sequence for increasing $M$ is not exactly the same as the
sequence in terms of distances from the sun.
For example, the positions of Venus and Mercury are reversed and we
will see later that our
formula indicates 5 objects between Mars and Jupiter---these
could be asteroids and Pluto---and
one between Jupiter and Saturn (another asteroid).

The table below gives $M$,
and its offset value $Q = M+0.337$, for
the planets in order of increasing values are:

\vspace{2.5mm}

\begin{tabular}{||r|r|r||} \hline
planet & $M = \frac{d}{p}$ & Q = M + 0.337 \\ \hline
Venus & $-2.98 \times 10^{-3}$ & 0.334 \\
Mercury & $6.6 \times 10^{-3}$ & 0.343 \\
Earth & 1.00 & 1.337 \\
Mars & 1.487 & 1.824 \\
Pluto & 6.17 & 6.507 \\ 
Jupiter & 12.5 & 12.837 \\
Saturn & 21.53 & 21.867 \\
Uranus & 29.51 & 29.847 \\
Neptune & 39.04 & 39.377 \\ \hline
\end{tabular}

\vspace{2.5mm}

The $Q= \frac{d}{p} + 0.337$ values of the major asteroids are:

\begin{quote}
Ceres (7.16), Pallas (8.86), Juno (9.17), Vesta (11.02),
Astraea (4.0), Hebe (8.326), Iris (8.38), Hygiea (4.42),
Eunomia (10.76), Euphrosyne (14.08).
\end{quote}

These numbers and those in the Table above have been computed
from the information in Reference 1.

We now propose the following formula
for the Q-numbers for the planets:

\begin{equation}
Q (n+1) = 1.361 Q (n)
\end{equation}
where $Q (0) = 0.2863$. This means that the period, $p(n)$ in days, of
the $n$th planet is given by:

\begin{equation}
p (n) = \frac{d(n)}{(1.361)^n \times 0.2863 - 0.337}
\end{equation}

Here the sequence order is Venus (0), Mercury (1), 
Earth (5), Mars (6), Jupiter (12), Saturn (14), Uranus (15),
and Neptune (16).

Given 17 items, one can interpolate by using a
polynomial that is of 16th degree, with 17 constants.
Our formula uses just 3 constants and the $Q$-values are
fixed by only 2 numbers.
So our formula provides significant compression of
information.
An interesting question to ask is: What is the best that can
be done in terms of such compression?

Beginning with the $Q-$number for earth
in this sequence, namely $1.337$, and multiplying successively by  
1.361, We get the numbers:

\[1.337, 1.820, 2.477, 3.371, 4.587, 6.243, 8.497, 11.565, 15.740, 21.422, 29.155, 39.680\]

We obtain good agreement with the value for Mars (1.820), which is followed
by 5 additional values (these could be asteroids and Pluto),
the value for Jupiter (11.565) which is off by about 10 percent,
another value (15.74) that is near the correct value for
the asteroid Euphrosyne (14.08), and then very good agreement with the
correct values for Saturn, Uranus, and Neptune.

Considering the values intermediate to those for Mars and Saturn, we have:
2.477 (no asteroid known to the author); 3.371 (Astraea);
4.587 (Hygiea); 6.243 (Ceres; but Pluto is closer);
8.497 (Hebe, Iris, Pallas); 11.565 (Vesta, and Jupiter);
15.74 (Euphrosyne).

On the other hand, by dividing 1.337 successively by 1.361, we get:

\[0.982, 0.722, 0.530, 0.390, 0.286\]

There are no objects for the first three of these values;
the next two, come closest to the values for Mercury
and Venus. 
When the constant 0.337 is subtracted, the $M$ values for
Mercury and Venus obtained from the formula are
$5.4 \times 10^{-2}$ and 
$-5.0 \times 10^{-2}$.
These are about one order of magnitude off, but provide the
correct direction of rotation for Venus.

When we consider the above information in conjunction with the
Titius-Bode law, it becomes clear why there could not
be a single planet at the distance of 2.8 AU, because our
rotation formula calls for 5 objects between Mars and Jupiter.

The extra three inferior planets provided by the formula
with the values of $Q = 0.532, 0.723, 0.983$ may have
been captured by the earth and Mars as their
satellites: Moon, Phobos, Deimos. 

Our formula (1) depends on just 2 constants:
the multiplication factor of 1.361 and the
starting value for $Q(0) = 0.2863$.
That just two constants are able to provide
a good fit for the rotation periods of most of the planets and many 
asteroids suggests that this formula may be
of some relevance in representing the dynamics of
the early solar system.

\section*{Reference}

\begin{enumerate}

\item V. Illingworth (ed.), {\it Dictionary of Astronomy.}
(Facts on File, New York, 1994)
\end{enumerate}
\end{document}